# Searching for Biophysically Realistic Parameters for Dynamic Neuron Models by Genetic Algorithms from Calcium Imaging Recording

**Magdalena Fuchs**
TU Wien
Austria

**Manuel Zimmer**
Institute of
Molecular Pathology
(IMP)
Vienna, Austria

**Radu Grosu**
TU Wien
Austria

**Ramin M. Hasani**[*]
TU Wien
Austria

## Abstract

Individual Neurons in the nervous systems exploit various dynamics. To capture these dynamics for single neurons, we tune the parameters of an electrophysiological model of nerve cells, to fit experimental data obtained by calcium imaging. A search for the biophysical parameters of this model is performed by means of a genetic algorithm, where the model neuron is exposed to a predefined input current representing overall inputs from other parts of the nervous system. The algorithm is then constrained for keeping the ion-channel currents within reasonable ranges, while producing the best fit to a calcium imaging time series of the AVA interneuron, from the brain of the soil-worm, *C. elegans* . Our settings enable us to project a set of biophysical parameters to the the neuron kinetics observed in neuronal imaging.

## Introduction

Even in a small organism such as the nematode *C. elegans* , individual neurons behave differently [1]. In order to map the complex behaviors such as learning [2], locomotion [3] and decision making [4] expressed by the nervous system, to the dynamics of the neuronal circuits, it is important to have detailed models of neurons with biologically realistic parameters. As a matter of fact it is not feasible to perform direct measurements of all the electrophysiological properties of ion channels and properties of individual neurons, even in smaller organisms such as *C. elegans* [1, 5, 6]. Therefore, predictions of neurons' variables by means of biophysically plausible mathematical models substantially help us to decode the mechanisms underlying behavior in the brain.

For ion channels within nerve cells [7, 8] or other types of cells [9, 10], parameters such as activation potentials and overall conductances can be measured by electrophysiological experiments. However, such experiments, have not been constructed for all types of neurons. Other parameters such as the ones describing intracellular calcium dynamics can not be easily determined from electrophysiological data. These parameters are nonetheless decisive for the exploitation of the overall behavior of neurons. With the rise of calcium imaging in neurobiology, reliable data for individual neuron types, has become available, including activity of entire neuronal networks [1]. Ergo, it is desirable to have a method to fit model's parameters for individual neurons by employing global brain imaging records. Genetic optimization methods have been used for fitting real data with simple neuron models within a network[11], for fitting detailed ion channel models [12] and complex multicompartmental models [13]. Here we show how a genetic algorithm (GA) can find biophysically meaningful parameters for a single neuron, by fitting the model-neuron calcium response to its corresponding experimentally

---

obtained calcium trace. In this way, we reason about the spontaneous dynamic activities of the neuron by an exclusive set of its ion channels' parameters.

**Calcium imaging revisit**

Calcium imaging is a key technique in modern biology in which a fluorescent calcium indicator is introduced to neurons of a transgenic animal, so that calcium concentrations inside the neuron can be measured as changes in fluorescence intensity when the organism is exposed to fluorescent-light [14]. In the dataset we use, the genetically encoded calcium indicator, CaMPK5, which was expressed in pan-neuronal fashion, and localized to the cell nuclei for whole brain imaging recordings [1].

The relation of intracellular calcium concentration and fluorescence intensity and and the temporal dynamics of this indicator are well-characterized and described in [15].

**Mapping calcium dynamics to the calcium indicator**

The electro-physiological neuron model computes a rough estimation of the concentration of the intracellular calcium. Calcium imaging data, as discussed, however, indicates a measured fluorescence intensity representing a relational projection to the inner calcium concentration. In order to link these together, we applied two modeling strategies;

Time-dynamics of the propagation of the intracellular calcium concentration to the nucleus, where the the fluorescent protein is located, were modeled by convolving the calcium trace with a Gaussian distribution of the with $\sigma$ = 0.9 s. The binding of the calcium to the GCaMP- protein was modeled by a Hill equation, as

$$\frac{\Delta F}{F} \propto d_{[Ca^{2+}]} = \frac{[Ca^{2+}]_{nuc}^n}{[Ca^{2+}]_{nuc}^n + k_d} \tag{1}$$

For GCaMP5K the half-activation $K_d$ = 0.189 mM and the Hill coefficient n = 3.8 [15].

**Neuron modeling framework revisit**

We use the SIM-CE [16] simulation platform for the model of the neuron, which is a single compartmental conductance-based model taking into account dynamics of voltage-gated calcium channels, voltage-gated potassium channels, calcium-gated potassium channels and a leak channel [17]. The activation of the voltage-gated channels is realized by a sigmoid shape, with temporal behavior represented as a linear first order differential equation. The calcium channels additionally, have a voltage-dependent inactivation mechanism with a longer time-constant, as suggested in [9]. Channel kinetics were simplified compared to [16], in order to reduce the number of parameters.

Intracellular calcium dynamics are described by first-order kinetics binding and de-binding to a storage molecule and transport out of the cell via a calcium-dependent pump, as described in [18] which undergoes a calcium-dependent increase in pump conductance on a longer timescale.

The model is implemented in MATLAB. Differential equations in the model were solved by using the built-in `ode23s` solver, a modified Rosenbrock method [19].

# Experiment

Figure 1A represents the calcium imaging trace for the AVA interneuron, known as a possible modulator for the reversal movement. Cell is getting active with various amplitudes, observed to be proportional to the duration on which it has been at its resting potential (1) and (2). There is a nonlinear decrease over some of the activation plateaus shown by arrow (3). This exponential decay continues over the immediate upcoming activation phases (4). Additionally, we observe that fall-phases, do not necessarily bring the neuron to the resting value (5). To capture such dynamics and reason about their biophysical causes, we apply a genetic algorithm to find the parameters for the neuron model described above.

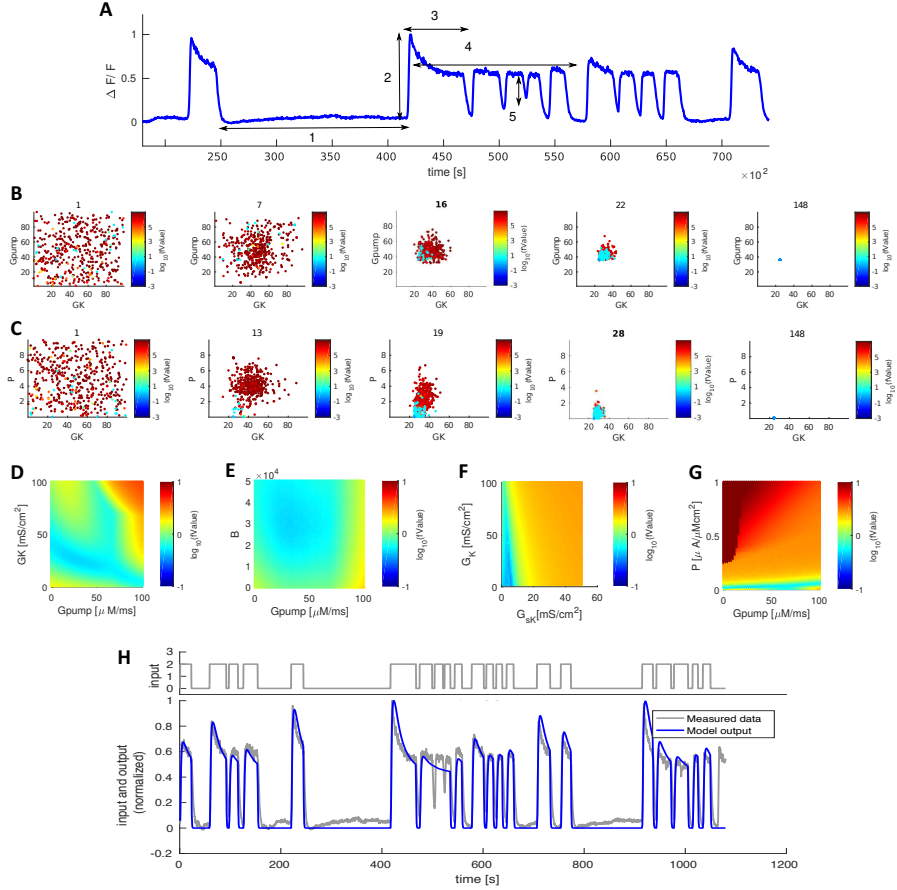

Figure 1: Optimization of a neuron-model's dynamics to fit real calcium traces, with a genetic algoritm. A) AVA interneuorn calcium imaging record (taken from [1]). B) Progressive representation of the GA for two parameters potassium-channel conductance $G_K$, and calcium pump, $G_{pump}$, in various generations. C) Progressive representation of the GA for two parameters potassium-channel conductance $G_K$, and calcium-channel conductivity factor, $P$, in various generations. D-G) Visualization of the impact of key pair-parameters' variation, for generating the correct AVA behavior, keeping other parameters fixed. H) Response of the neuron model (Blue line) plotted over the real AVA calcium trace (Grey line). Note that the model neuron is exposed to an input signal standing for net-input current to the nerve cell form other network components.

**Genetic algorithm revisit**

Genetic algorithms are a widely used global optimization method, which can be applied to a wide range of problems. [20]. They find the global optimum of a function of multiple parameters, called the cost function in a parameter space by calculating the cost function for multiple sets of parameters. Out of these parametersets, the ones with the best cost function values are picked out and modified in a random fashion, in order to generate a new "generation" of parametersets, with a better cost function. This is repeated until a satisfactory solution is found.

**GA Optimization of the AVA Neuron**

The model is composed of 35 parameters. This set includes channel conductances, channel activation and inactivation voltages and half-widths, parameters describing binding and de-binding of calcium in the cell soma and the activity and time-dynamics of the calcium pump. Variation of some parameters is set based on the available experimental reports [8, 21]. For parameters, where there is no experimental data available, the variation is selected in a range 0.1 to 10 times a center value, reported in other modeling studies [22, 23, 17, 24].

Input current to the model-cell, is generated from the calcium traces' rising edges and falling edges. In order to restrict the number of variables, currents were box-shaped current-pulses with an amplitude of 2 $\mu Acm^{-2}$.

The genetic algorithm optimization is undertaken using the MATLAB Genetic Algorithm Toolbox. We deployed an intermediate crossover and adaptive-feasible mutation function. The crossover fraction was set to 0.5, the number of individuals, that are transferred into the next generation without change was set zo 5, and the population size was selected to be 500.

Projections of all the parameter-space are plotted during optimization in order to capture all information for optimal starting points for further runs of the optimization (A representation of this step is shown in Figure 1B and 1C). The initial population was partially created by values obtained from reported experiments [16] and partially randomly sampled between the parameter boundaries.

The **cost function** was selected as the mean-squared difference between the normalized "[Ca]-bound-to-G-CaMP"-curve ($\frac{\Delta F}{F}$), and the measured fluorescence curve $d_{Ca}$, with penalties for currents exceeding a biologically realistic treshold $< p_I{}^2 >$, and intracellular calcium below a minimum concentration $< p_{[Ca]}{}^2 >$, as follows:

$$f = 30 \cdot < \frac{\Delta F}{F} - d_{[Ca^{2+}]} >^2 + \sum p_I I + 10 \cdot p_{Ca};$$

with

$$p_I = \begin{cases} 0 & |I| < I_{max} \\ |I| - I_{max} & |I| > I_{max} \end{cases}$$

$$p_{Ca} = \begin{cases} 0 & |[Ca]| > [Ca]_{min} \\ |[Ca]| - [Ca]_{min} & |[Ca]| < [Ca]_{min} \end{cases}$$

In order to understand the progression of the cost function with changes in key model parameters, we calculated cost function values on a grid of different parameter values. As expected, parameters such as the two potassium conductances are inversely correlated (Figure 1F). It is also plausible, that the cost function increases sharply with increasing calcium conductance, (Figure 1G), and that the amount of calcium storage molecules B has less influence. Calcium pump, $G_pump$ realizes stable dynamics, given a proper potassium, $G_K$ and calcium conductances, $p$. Nonetheless, Figure 1D shows, $G_K$ and $G_Pump$ depend on each other in a highly nonlinear fashion.

## Discussions

Biophysically plausible parameters were found for the model. Many dynamics of the AVA neuron were captured by the GA optimization process (i.e. properties 1, 2, 3 and 4. introduced in Figure 1A). Although the performance of the model is reasonably high, it has misinterpreted property 5 shown in Figure 1A.

Moreover, since the cost function can be arbitrarily chosen, various information about the desired output such as constraints on the current magnitudes can easily be incorporated into the optimization platform. Compared to models without plausible biological properties, over-fitting is not an issue, since the model, by design, will be able to produce realistic traces, even if over-trained for a single trace. Contribution of single parameters have to be assessed carefully and a minimum of necessary parameters should be chosen. If this is not done, biological relevance might be interpreted into parameter values, which, in fact, do not have a large impact on the model's output. In order to obtain the exact values for the parameters, once the curve is close to the desired output value, consecutive employment of a gradient-based optimization method, might provide fairly better results which will be the focus of our future investigations.

The platform enables us to compare biophysical properties of neurons with different dynamics and further assists understanding of how behavior emerges from multi-dimensional neuron dynamics in the nervous system.

## Acknowledgments

Authors would like to thank Microsoft Azure for providing computational resources through the Microsoft Azure Award for Research Program.

## References

[1] Saul Kato, Harris S Kaplan, Tina Schrodel, Susanne Skora, Theodore H Lindsay, Eviatar Yemini, Shawn Lockery, and Manuel Zimmer. Global brain dynamics embed the motor command sequence of Caenorhabditis elegans. *Cell*, 163(3):656–669, 2015.

[2] Evan L Ardiel and Catharine H Rankin. An elegant mind: learning and memory in Caenorhabditis elegans. *Learning & Memory*, 17(4):191–201, 2010.

[3] Stephen R Wicks and Catharine H Rankin. Effects of tap withdrawal response habituation on other withdrawal behaviors: the localization of habituation in the nematode Caenorhabditis elegans. *Behavioral neuroscience*, 111(2):342, 1997.

[4] Evan Z Macosko, Navin Pokala, Evan H Feinberg, Sreekanth H Chalasani, Rebecca A Butcher, Jon Clardy, and Cornelia I Bargmann. A hub-and-spoke circuit drives pheromone attraction and social behavior in C. elegans. *Nature*, 458(7242):1171, 2009.

[5] William M Roberts, Steven B Augustine, Kristy J Lawton, Theodore H Lindsay, Tod R Thiele, Eduardo J Izquierdo, Serge Faumont, Rebecca A Lindsay, Matthew Cale Britton, Navin Pokala, et al. A stochastic neuronal model predicts random search behaviors at multiple spatial scales in C. elegans. *Elife*, 5:e12572, 2016.

[6] Annika LA Nichols, Tomáš Eichler, Richard Latham, and Manuel Zimmer. A global brain state underlies C. elegans sleep behavior. *Science*, 356(6344):eaam6851, 2017.

[7] Hiroshi Suzuki, Rex Kerr, Laura Bianchi, Christian Frøkjær-Jensen, Dan Slone, Jian Xue, Beate Gerstbrein, Monica Driscoll, and William R Schafer. In vivo imaging of C. elegans mechanosensory neurons demonstrates a specific role for the mec-4 channel in the process of gentle touch sensation. *Neuron*, 39(6):1005–1017, 2003.

[8] Shi-Qing Cai, Yi Wang, Ki Ho Park, Xin Tong, Zui Pan, and Federico Sesti. Autophosphorylation of a voltage-gated k+ channel controls non-associative learning. *The EMBO journal*, 28(11):1601–1611, 2009.

[9] Maëlle Jospin, Vincent Jacquemond, Marie-Christine Mariol, Laurent Ségalat, and Bruno Allard. The L-type voltage-dependent Ca2+ channel EGL-19 controls body wall muscle function in caenorhabditis elegans. *The Journal of cell biology*, 159(2):337–348, 2002.

[10] Shangbang Gao and Mei Zhen. Action potentials drive body wall muscle contractions in Caenorhabditis elegans. *Proceedings of the National Academy of Sciences*, 108(6):2557–2562, 2011.

[11] Jörn Fischer, Poramate Manoonpong, and S Lackner. Reconstructing neural parameters and synapses of arbitrary interconnected neurons from their simulated spiking activity. *arXiv preprint arXiv:1608.06132*, 2016.

[12] Meron Gurkiewicz and Alon Korngreen. A numerical approach to ion channel modelling using whole-cell voltage-clamp recordings and a genetic algorithm. *PLOS Computational Biology*, 3(8):1–15, 08 2007.

[13] Pablo Achard and Erik De Schutter. Complex parameter landscape for a complex neuron model. *PLoS Computational Biology*, 2(7):e94, 2006.

[14] Christine Grienberger and Arthur Konnerth. Imaging calcium in neurons. *Neuron*, 73(5):862–885, 2012.

[15] Jasper Akerboom, Tsai-Wen Chen, Trevor J Wardill, Lin Tian, Jonathan S Marvin, Sevinç Mutlu, Nicole Carreras Calderón, Federico Esposti, Bart G Borghuis, Xiaonan Richard Sun, et al. Optimization of a gcamp calcium indicator for neural activity imaging. *Journal of Neuroscience*, 32(40):13819–13840, 2012.

[16] Ramin M. Hasani, Victoria Beneder, Magdalena Fuchs, David Lung, and Radu Grosu. SIM-CE: An advanced simulink platform for studying the brain of Caenorhabditis elegans. *arXiv preprint arXiv:1703.06270*, 2017.

[17] Masahiro Kuramochi and Yuishi Iwasaki. Quantitative modeling of neuronal dynamics in C. elegans. In *Neural Information Processing. Theory and Algorithms*, pages 17–24. Springer, 2010.

[18] David Sterratt, Bruce Graham, Andrew Gillies, and David Willshaw. *Principles of computational modelling in neuroscience*. Cambridge University Press, 2011.

[19] Lawrence F Shampine and Mark W Reichelt. The matlab ode suite. *SIAM journal on scientific computing*, 18(1):1–22, 1997.

[20] SN Sivanandam and SN Deepa. *Introduction to genetic algorithms*. Springer Science & Business Media, 2007.

[21] Viviane Lainé, Christian Frøkjær-Jensen, Harold Couchoux, and Maëlle Jospin. The $\alpha 1$ subunit egl-19, the $\alpha 2/\delta$ subunit unc-36, and the $\beta$ subunit ccb-1 underlie voltage-dependent calcium currents in caenorhabditis elegans striated muscle. *Journal of Biological Chemistry*, 286(42):36180–36187, 2011.

[22] Ramin M Hasani, Magdalena Fuchs, Victoria Beneder, and Radu Grosu. Non-associative learning representation in the nervous system of the nematode Caenorhabditis elegans. *arXiv preprint arXiv:1703.06264*, 2017.

[23] T Aoyagi, Y Kang, N Terada, T Kaneko, and T Fukai. The role of ca 2+-dependent cationic current in generating gamma frequency rhythmic bursts: modeling study. *Neuroscience*, 115(4):1127–1138, 2002.

[24] Stephen R Wicks, Chris J Roehrig, and Catharine H Rankin. A dynamic network simulation of the nematode tap withdrawal circuit: predictions concerning synaptic function using behavioral criteria. *The Journal of Neuroscience*, 16(12):4017–4031, 1996.